\documentclass[11pt,preprintnumbers,aps,amssymb,nofootinbib,amsmath,superscriptaddress,notitlepage,prd]
{revtex4-2}
\usepackage{graphicx}
\usepackage{bm} 
\usepackage{color} 
\usepackage{hyperref}
\usepackage{tensor} 
\usepackage{color} 
\usepackage{slashed} 
\usepackage{relsize}	
\usepackage{soul} 
\usepackage{hyperref}
\usepackage{tensor} 
\usepackage{yfonts} 
\def\bal#1\gal{\begin{align}#1\end{align}}

\newcommand{\eq}[1]{(\ref{#1})}
\newcommand{\fig}[1]{Fig.~\ref{#1}}
\renewcommand{\sec}[1]{Sec.~\ref{#1}}
\newcommand{\Ai}{\text{Ai}}

\begin{document}
\title{Sokolov--Ternov effect in rotating systems}
\author{Jonathan D. Kroth}
\affiliation{Department of Physics and Astronomy, Iowa State University, Ames, Iowa, 50011, USA}

\author{Kirill Tuchin}
\affiliation{Department of Physics and Astronomy, Iowa State University, Ames, Iowa, 50011, USA}
\date{\today}

\begin{abstract}

We study electromagnetic radiation by electrically charged fermions embedded in a rotating medium in an external magnetic field. We compute the dependence of the radiation intensity on the angular velocity $\Omega$ of the rotating medium for fermion polarizations along and opposite the magnetic field direction in intense subcritical fields. The polarization dependence of the photon radiation results in the Sokolov--Ternov effect---the radiative polarization of fermions. We study the dependence of the degree of polarziation on $\Omega$. We found that rotation significantly changes the degree of polarization. We show that the rotating quark-gluon plasma acquires a finite magnetic moment that exhibits complex dependence on $\Omega$.

\end{abstract}

\maketitle
\section{Introduction}\label{sec:intro}

The electromagnetic radiation by a charged fermion moving in an external magnetic field depends on the polarization of the fermion. This effect was first proposed by Sokolov and Ternov  \cite{sokolov1963sokolov}, who computed the photon radiation rate for different polarization states of the fermion using exact solutions of the Dirac equation in a constant magnetic field and leading order perturbation theory for the photon emission. The Sokolov--Ternov effect is the self-polarization of relativistic fermions moving in a magnetic field due to spin-flip transitions. It has been observed in storage rings of high energy colliders \cite{SokolovAndTernov}. A detailed theory of synchrotron radiation using the Sokolov--Ternov approach is presented in \cite{book:Bordovitsyn}. That theory was generalized in \cite{Derbenev:1973ia,Derbenev:1972mk,Bell:1986ir,Barber:1988zz} to include recoil effects, see a review \cite{Mane:2005xh} and references therein. The focus of the above-cited work is the polarization of fermions in storage rings where the magnetic field strength is relatively weak. In tabletop accelerator proposals, charged fermions are accelerated using the plasma wake fields produced by high-intensity laser beams. In such designs the fields are very strong, approaching Schwinger's critical value. The radiative electron polarization by ultrafast laser pulses is discussed in  \cite{DelSorbo:2017fod,Seipt:2018adi}. The radiative fermion polarization in the strong field regime is reviewed in  \cite{Gonoskov:2021hwf}. The radiative polarization should not be confused with the Barnett effect, which is the induction of non-zero magnetization in a rotating body \cite{barnett1915,barnett1935,Buzzegoli:2022qrr}.

This work considers the Sokolov--Ternov effect in a different physical context. Our main motivation is to investigate the properties of quark-gluon plasma, produced in relativistic heavy-ion collisions. Relativistic heavy-ion collisions are characterized by very intense, though subcritical, magnetic fields \cite{Kharzeev:2007jp,Skokov:2009qp,Voronyuk:2011jd,Bzdak:2011yy,Bloczynski:2012en,Deng:2012pc,Tuchin:2013apa,Zakharov:2014dia} and extremely high vorticity \cite{STAR:2017ckg,Csernai:2013bqa,Deng:2016gyh,Jiang:2016woz,Xia:2018tes,Becattini:2015ska}. Due to the axial symmetry of the colliding system, its average vorticity and magnetic field are either parallel or anti-parallel. The magnetic field and vorticity affect the fermion's motion in the transverse plane. There are two relevant length scales: the magnetic length $1/\sqrt{qB}$, which sets the transverse correlation length in a magnetic field, and $1/\Omega$, which is the location of the light cylinder---the causal boundary of a rotating system. Throughout the paper we use the natural units $\hbar=c=1$ and set the energy scale to be the fermion mass: $M=1$.

At the early stages of a heavy-ion collision, $qB\gg \Omega^2$, which in \cite{Buzzegoli:2022dhw,Buzzegoli:2023vne} was referred to as ``relatively slow" rotation, even though the absolute value of the angular velocity is the largest in any known physical system. In this regime, one can ignore the light-cone boundary, owing to the exponential decay of the fermion wave functions at radial distances greater than the magnetic length. The corresponding fermion states were derived in \cite{Buzzegoli:2022dhw,Buzzegoli:2023vne}, where we computed the intensity of the synchrotron radiation by a fermion that is dragged along with the plasma in rotating motion with angular velocity $\Omega$. In this work, we apply the results of \cite{Buzzegoli:2022dhw,Buzzegoli:2023vne} to compute the effect of rotation on the radiative fermion polarization. At the later stages of a relativistic heavy-ion collision, $qB\ll \Omega^2$, and one has to take account of causality by restricting the fermion current to the interior of the light cylinder. The electromagnetic radiation is quite different in this regime \cite{Buzzegoli:2024nzd}, and we defer its study to future work.

In this work, the calculation of the radiation rate and intensity is done using leading order perturbation theory. The fermion wavefunctions are obtained as exact solutions of the Dirac equation with uniform rotation and constant magnetic field. The photon wave functions are solutions to the wave equation and eigenfunctions of the curl operator. In the limit of vanishing angular velocity, this setup is identical to the original approach of Sokolov and Ternov. In view of phenomenological applications, we study intense subcritical magnetic fields  $|qB|\lesssim 1$. Namely, we chose $|qB|=10^{-3},10^{-2}$. The range of fermion energies is $E=1-15$. The corresponding values of the parameter $\chi=|eF_{\mu\nu}p^\nu|$ evaluated at $\Omega=0$ are $0.01<\chi<0.15$. The values of the angular velocity are $|\Omega|=10^{-6},10^{-5}$.

The paper is structured as follows. \sec{sec:review} is a review of the analytical results obtained in \cite{Buzzegoli:2022dhw,Buzzegoli:2023vne} with a focus on the fermion polarization states. Our numerical results for the rate and intensity of radiation for different fermion polarziation states are presented in \sec{sec:numerics}. The main result is a strong dependence of the degree of polarization $\eta$ on the magnitude and direction of the angular velocity. \fig{fig:STRatio} presents $\eta$ as a function of $\chi$.  The radiative polarization leads to magnetization of the rotating system.  The degree of magnetization is shown in \fig{fig:qgp-mag}.

\section{Sokolov-Ternov formalism for rotating fermions}\label{sec:review}

\subsection{Fermion Wavefunction}\label{sec:fermion-WF}

Consider a particle with charge $q$ and mass $M$ subject to a constant magnetic field $\bm{B} = B \bm{\hat{z}}$, with $B > 0$ and $qB = -|qB|$. Consider further that the particle is embedded in a medium which rotates with fixed angular velocity $\bm{\Omega} = \Omega \bm{\hat{z}}$. By embedded, we mean that centrifugal forces do not cause the particle to be ejected from the system. To mathematically realize this, we place ourselves in the rest frame of the medium, then rotate with angular velocity $-\bm{\Omega}$ into a ``lab" frame and proceed  in cylindrical coordinates $\{t,r,\phi,z\}$. 
The Dirac equation for such a particle can be written in the Schr\"odinger form $i \partial_t \psi = \hat{H} \psi$, with
\begin{equation}
    \hat{H} = \gamma^0 \bm{\gamma} \boldsymbol{\cdot} (\bm{p} - q\bm{A}) + \gamma^0 1 + \Omega \hat{J}_z ,
    \label{eq:Hamiltonian}
\end{equation}
where $\bm{p} = -i \nabla$ and $\hat{J}_z = -i\partial_\phi + \frac{i}{2} \gamma^x \gamma^y$. We work in the symmetrical gauge $A_\mu = \left(0,-By/2,Bx/2,0\right)$. This is the ordinary Dirac Hamiltonian with the additional term $\bm{\Omega} \boldsymbol{\cdot} \bm{J}$, which accounts for the rotation. In this form we see clearly that $\hat{H}$ commutes with $\hat{p}_z$ and $\hat{J}_z$. Denote their corresponding eigenvalues by $E$, $p_z$, and $m$.
The eigenstates of the Hamiltonian (\ref{eq:Hamiltonian}) without rotation in the Dirac representation of the $\gamma$-matrices is well-known \cite{SokolovAndTernov}. We get the solution for the rotating particle by adding $m \Omega$ to $E$.

The energy eigenvalues $E$ are
\begin{equation}\label{LL}
    E = \sqrt{1 + 2n |qB| + p_z^2} + m \Omega
\end{equation}
where $n$ is a non-negative integer. These are ordinary Landau levels with an additional term due to the rotation. The radial dependence of the wavefunction is generally a Laguerre function,
\begin{equation}
    I_{n,a}(x) = \sqrt{\frac{a!}{n!}} e^{-x/2} x^{(n-a)/2} L_{a}^{(n-a)}(x)
\end{equation}
where $L_a^{(\alpha)}(x)$ is a generalized Laguerre polynomial. The fermion wavefunction is then
\begin{equation}
    \psi(t,r,\phi,z) = e^{-iEt} \frac{e^{i p_z z} e^{i m \phi}}{\sqrt{L} \sqrt{2\pi}}
    \begin{pmatrix}
        C_1 I_{n-1,a}(\rho) e^{-i\phi/2} \\
        i C_2 I_{n,a}(\rho) e^{i\phi/2} \\
        C_3 I_{n-1,a}(\rho) e^{-i\phi/2} \\
        i C_4 I_{n,a}(\rho) e^{i\phi/2}
    \end{pmatrix}
    \label{eq:FermionWavefunction}
\end{equation}
 where $\rho = \frac{|qB|}{2} r^2$ and $a$ is an auxiliary quantum number defined by $m = n - a - \frac{1}{2}$; $a$ takes on non-negative integer values.
 The coefficients $C_i$ can be chosen so that $\psi$ is also an eigenstate of the magnetic moment $\mu_z$ with eigenvalue $\zeta$:
\begin{gather}
    \bm{\mu} = \bm{\Sigma} - \frac{i \gamma^0 \gamma^5}{2} \bm{\Sigma} \times (\bm{p} - q \bm{A}), \\
    \mu_z \psi = \zeta \sqrt{(E-m\Omega)^2 - p_z^2} \psi .
\end{gather}
 $\zeta$ thus indicates the transverse (magnetic) polarization of the fermion's spin. Along with the normalization
\begin{equation}
    \int  \psi^\dagger \psi \, d^3x = 1,
\end{equation}
this fixes the coefficients $C_i$:
\begin{gather}
    C_{1,3} = \frac{1}{2\sqrt{2}} B_+ (A_+ \pm \zeta A_-) , \qquad C_{2,4} = \frac{1}{2\sqrt{2}} B_- (A_- \mp \zeta A_+) , 
    \label{eq:CDef1} \\
    A_{\pm} = \left(1 \pm \frac{p_z}{E-m\Omega}\right)^{1/2} , \qquad B_{\pm} = \left(1 \pm \frac{\zeta}{\sqrt{(E-m\Omega)^2 - p_z^2}}\right)^{1/2} .
    \label{eq:CDef2}
\end{gather}

The rotation of the system imposes a causal boundary: the ``velocity" of the system at radius $r$ is $r|\Omega|$, which cannot exceed $c = 1$. This imposes an upper cutoff on the allowed values of $n$ and $a$. For example, we can compute $\langle r^2 \rangle$:
\begin{equation}
    \langle r^2 \rangle = \int  \psi^\dagger_{n,a,p_z,\zeta} r^2 \psi_{n,a,p_z,\zeta} \, d^3x = \frac{2}{|qB|} \left[(n+a+\frac{1}{2}) - \langle S_z \rangle\right]
\end{equation}
which implies that 
\begin{equation}
    n,a \ll N_{\text{caus}} = \frac{|qB|}{2\Omega^2} .
    \label{eq:Causality}
\end{equation}
In the regime $|qB| \gg \Omega^2$, physically meaningful results should not depend on $N_\text{caus}$. Imposing (\ref{eq:Causality}) ensures that $\psi$ is exponentially suppressed as we approach the causal boundary, so we can extend any integrations over $r$ to infinity without affecting our results. We explicitly demonstrated in \cite{Buzzegoli:2023vne} that the intensity of photon radiation does not depend on $N_\text{caus}$ in this limit. The same bound for $a$ can be found by considering the degeneracy of the Landau levels \cite{Chen:2015hfc}.

\subsection{Photon Wavefunction}\label{sec:photon-WF}
We choose to write our photon wavefunctions in cylindrical coordinates as well. However, we do not regard the photons as rotating with the fermion in the medium, since the interactions of photons with quark-gluon plasma are weak. In the lab frame, the radiated photon wave function is a solution to the wave equation in an inertial frame. We work in the radiation gauge, $A^0 = 0$, $\bm{\nabla} \boldsymbol{\cdot} \bm{A} = 0$, and choose wavefunctions which are eigenfunctions of the curl operator:\footnote{We have absorbed a factor of $k/k_\perp$ used in our previous work \cite{Buzzegoli:2022dhw,Buzzegoli:2023vne}  into the definitions of $\bm{T}$ and $\bm{P}$.}
\begin{gather}
    \bm{A}_{h,\ell,k_\perp,k_z} = \frac{1}{\sqrt{2\omega V}} \bm{\Phi}_{h,\ell,k_\perp,k_z} e^{-i\omega t} \label{photon.w.f.}\\
    \bm{\Phi}_{h,\ell,k_\perp,k_z} = \frac{1}{\sqrt{2}} (h \bm{T}_{\ell,k_\perp,k_z} + \bm{P}_{\ell,k_\perp,k_z}) e^{ik_z z + i \ell \phi} \\
    \bm{T}_{\ell,k_\perp,k_z} = \frac{i\ell}{k_\perp r} J_\ell (k_\perp r) \bm{\hat{r}} - J'_\ell(k_\perp r) \bm{\hat{\phi}} \\
    \bm{P}_{\ell,k_\perp,k_z} = \frac{i k_z}{k} J'_\ell(k_\perp r) \bm{\hat{r}} - \frac{\ell k_z}{k k_\perp r} J_\ell(k_\perp r) \bm{\hat{\phi}} + \frac{k_\perp}{k} J_\ell(k_\perp r) \bm{\hat{z}} .
\end{gather}
$k = \omega$ is the photon energy, $\theta$ is the polar angle at which the photon is emitted, $k_\perp = \omega \sin\theta$ is the transverse momentum, $k_z = \omega \cos\theta$ is the longitudinal momentum, $\ell$ is the photon angular momentum, and $h = \pm 1$ is the photon helicity ($+1$ for right-handed, $-1$ for left-handed). $\bm{T}$ and $\bm{P}$ are called the toroidal and poloidal components of $\bm{A}$, respectively. $\bm{A}_{h,\ell,k_\perp,k_z}$ is normalized to one photon per unit volume. A general photon wavefunction is a linear combination of these basis functions.

\subsection{Differential Radiation Rate and Intensity}\label{sec:rate}

The rate $\dot w$ and intensity $W$ of electromagnetic radiation can be computed using the exact wavefunctions  \eq{eq:FermionWavefunction} and \eq{photon.w.f.} in leading order perturbation theory. We compute the radiation intensity to compare with our previous results and the rate to compute the degree of polarization due to the Sokolov--Ternov effect. The result takes the simplest form in the frame where  $p_z = 0$ \cite{Buzzegoli:2023vne}:
\begin{align}
    \dot{w}^{n,a}_{\zeta,\zeta',h} &= \frac{q^2}{4\pi} \sum_{n',a'} \int_0^\pi d\theta \frac{\omega_0 \sin\theta}{1 + \frac{\omega_0 \cos^2\theta}{E'-m'\Omega}} |\langle \bm j \cdot \bm\Phi \rangle|^2 ,
    \label{eq:Rate} \\
    W^{n,a}_{\zeta,\zeta',h} &= \frac{q^2}{4\pi} \sum_{n',a'} \int_0^\pi d\theta \frac{\omega_0^2 \sin\theta}{1 + \frac{\omega_0 \cos^2\theta}{E'-m'\Omega}} |\langle \bm j \cdot \bm\Phi \rangle|^2 ,
    \label{eq:RadInt}
\end{align}
where we explicitly indicated the input quantum numbers on the left-hand sides. Primed quantities refer to the final fermion state. In contrast to our previous work, in this paper we are interested in the spin- and helicity-dependence of the radiation, so we do not sum these quantum numbers. The photon energy $\omega_0$ follows from the energy and longitudinal momentum conservation:
\begin{align}
    \omega_0 &= \frac{E-m'\Omega}{\sin^2\theta} \left(1 - \left[1 - \frac{\mathcal{B} \sin^2\theta}{(E-m'\Omega)^2}\right]^{1/2}\right) \label{omega0}\,,\\
    \mathcal{B} &= 2(n-n')|qB| + (m-m')^2\Omega^2 + 2(E-m\Omega)(m-m')\Omega\, .
\end{align}
The transition amplitude reads:
\begin{equation}
    \begin{split}
        \langle \bm j \cdot \bm\Phi \rangle = \frac{1}{\sqrt{2}} I_{a,a'}(x) \big(&\sin\theta [K_4 I_{n-1,n'-1}(x) - K_3 I_{n,n'} (x)] \\
        &+ K_1 (h-\cos\theta) I_{n,n'-1}(x) - K_2 (h+\cos\theta) I_{n-1,n'}(x)\big) .
    \end{split}
    \label{eq:jPhi}
\end{equation}
The argument of the $I$-functions is
\begin{equation}
    x = \frac{\omega_0^2 \sin^2\theta}{2|qB|} .
\end{equation}
The $K_i$ are combinations of the $C_i$ defined in Eqs. (\ref{eq:FermionWavefunction}), (\ref{eq:CDef1}), and (\ref{eq:CDef2}) for the initial and final states:
\begin{equation}
    \begin{split}
        K_1 = C_1'C_4 + C_3'C_2 , \quad K_2 = C_4'C_1 + C_2'C_3 , \\
        K_3 = C_4'C_2 + C_2'C_4 , \quad K_4 = C_1'C_3 + C_3'C_1 .
    \end{split}
    \label{eq:KDef}
\end{equation}

\subsection{Reflection symmetry}
\label{sec:RefSym}

The system is symmetric under a reflection across the $xy$-plane, as we will now demonstrate. This can be deduced by analyzing each part of Eq.~(\ref{eq:jPhi}). The effect of this reflection on physical quantities is the following:
\begin{equation*}
    \left\{p_z , p_z' , \theta , h \right\} \to
    \left\{-p_z , -p_z' , \pi - \theta , -h \right\} .
\end{equation*}

Now analyze Eq.~(\ref{eq:jPhi}). Since $\omega_0$ and $x$ are unchanged, the $I$ functions are as well. The terms with $(h\pm\cos\theta)$ factors acquire minus signs. The $K_i$ can be transformed from their defintions in Eqs. (\ref{eq:CDef1}), (\ref{eq:CDef2}), and (\ref{eq:KDef}): $A_{\pm} \to A_{\mp}$ and $B_{\pm}$ is unchanged. So $C_{1/3} \to \pm \zeta C_{1/3}$ and $C_{2/4} \to -\zeta C_{2/4}$. thus $K_{1,2} \to \zeta \zeta' K_{1,2}$ and $K_{3,4} \to - \zeta \zeta' K_{3,4}$. So reflecting in the $xy$ plane gives $\langle \bm j \cdot \bm\Phi \rangle \to - \zeta \zeta' \langle \bm j \cdot \bm\Phi \rangle$, which is just a phase.

However, from this analysis we can observe that by reflecting only the momenta and the angle $\theta$, we accomplish the same (up to a phase) transformation as changing $h \to -h$. This observation is useful in the numerical analysis that we present in the next section: by computing the rate or intensity for a single photon helicity, we can get the rate or intentisy for the other helicity by changing $\theta \to \pi - \theta$. We also see that the two photon helicities contribute equally to the total rate and intensity, which are integrated over $\theta$.

\section{Numerical Results}\label{sec:numerics}

The rate (\ref{eq:Rate}) and intensity (\ref{eq:RadInt}) can only be computed numerically. We show the results of our numerical calculations here. In addition to these quantities, we also discuss the Sokolov--Ternov effect, i.e.\ the radiative self-polarization of electrons, and the effect that rotation has on this process. We compare our calculations to the semiclassical approximation derived by Sokolov and Ternov \cite{SokolovAndTernov}, which does not include the effects of rotation. This approximation is valid in the regime where $|qB|\ll 1$ and $E \gg 1$, which is satisfied for most of our parameters. The  parameter $\chi = |qB| Ev$ (evaluated at $\Omega=0$) does not exceed 0.15, implying that the recoil effect is small and the semiclassical approximation is in fact close to the classical expression. Nevertheless, since $\chi$ is not very small, we derive the semiclassical expressions for intensity and rate to all orders in $\chi$ in Appendix \ref{sec:Semiclassical}.

\subsection{Semiclassical limit for $\Omega=0$}

It is convenient to compare the results of our calculations with the classical results in the stationary case $\Omega=0$. These expressions, well-known in the literature, read \cite{SokolovAndTernov}:
\begin{align}
W_{\text{Cl}} &=\frac{2}{3} \alpha \chi^2 = \frac{2}{3} \alpha |qB|^2 E^2v^2\,,\label{Wclass}\\
 \dot{w}_{\text{Cl}} &= \frac{5 \sqrt{3}}{6} \alpha \frac{\chi}{E}= \frac{5 \sqrt{3}}{6} \alpha |qB|v \, ,\label{Wclass2}
\end{align}
where $v=\sqrt{1-1/E^2}$ is the incident fermion velocity. In the ultra-relativistic limit $v\to 1$ these expressions become
\begin{align}
W_{\text{Cl}}^* &= \frac{2}{3} \alpha |qB|^2 E^2\,,\label{Wclass*} \\
 \dot{w}_{\text{Cl}}^* &=\frac{5 \sqrt{3}}{6} \alpha |qB| \, . \label{Wclass2*}
\end{align}
In the following, we have normalized both our results for the numerically calculated rate and intensity, which include the effects of rotation, and the semiclassical rate and intensity, with $\Omega = 0$, by dividing by \eq{Wclass*} or \eq{Wclass2*}.
We found that, for $\Omega = 0$, the numerically calculated values and the semiclassical values were generally in good agreement. However, the two methods were found to disagree for the rates for processes which do not involve a spin flip. We corrected this by imposing an infrared cutoff on the integration over the emitted photon energy. The details of this cutoff and the semiclassical expansion can be found in Appendices \ref{sec:Semiclassical} and \ref{sec:IR}.

\subsection{Total intensity and rate}\label{sec:NumRate}

In our previous work \cite{Buzzegoli:2022dhw,Buzzegoli:2023vne}, we saw that rotation caused the radiation intensity to increase when $\bm{\Omega}$ and $q\bm{B}$ were antiparallel and decrease when they were parallel. This is generally reflected in our results here for both the rate and intensity. We can qualitatively understand this dependence by noting that fermion motion is a non-linear combination of two circular motions: one with angular velocity $\Omega$ due to the system rotation, and another with angular velocity (synchrotron frequency) $\omega_B = qB/E$ due to the magnetic field. When the direction of rotation due to the magnetic field coincides with the direction of the system rotation (e.g.\ $qB<0$ and $\Omega>0$), the two rotations ``add up" to enhance the radiation. Conversely, when the two rotations are in the opposite direction (e.g.\ $qB<0$ and $\Omega<0$), they ``subtract" to suppress the radiation. We also note that since the synchrotron frequency decreases as $E^{-1}$, the relative contribution of the magnetic field as compared with overall rotation decreases with $E$, meaning that the effect of $\bm{\Omega}$ (to enhance or suppress) increases.

\begin{figure}
    \centering
    \includegraphics[width=3in]{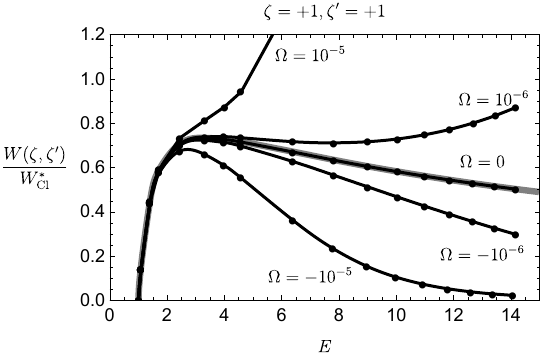} 
    \includegraphics[width=3in]{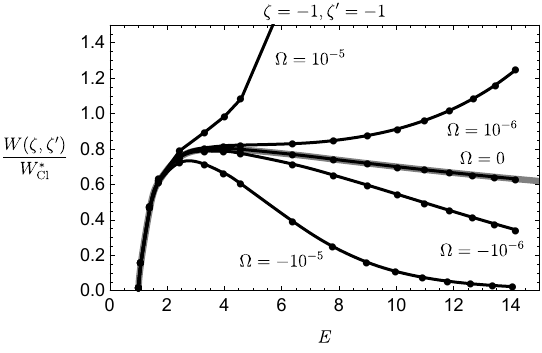} \\
    \includegraphics[width=3in]{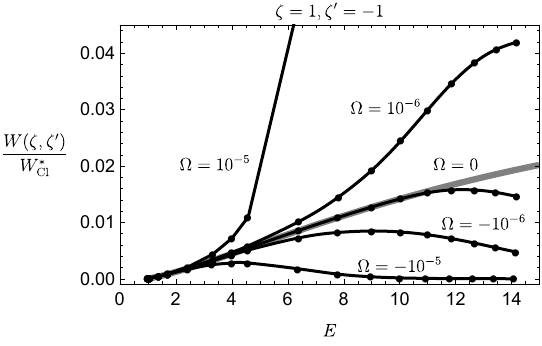} 
    \includegraphics[width=3in]{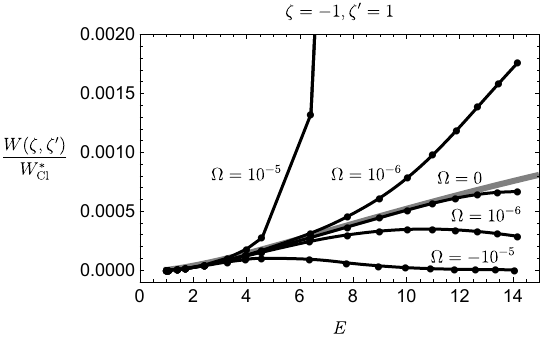} \\
    \caption{The total intensity for $qB = -10^{-2}$ and various $\Omega$. The gray line is the semiclassical result \eq{eq:SCInt} with the infrared cutoff (\ref{IRcutoff}). $W_{\text{Cl}}^*$ is given by \eq{Wclass*}.}
    \label{fig:TotIntB2}
\end{figure}

\begin{figure}
    \centering
    \includegraphics[width=3in]{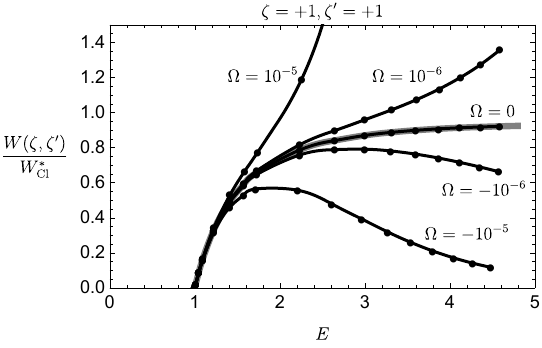} 
    \includegraphics[width=3in]{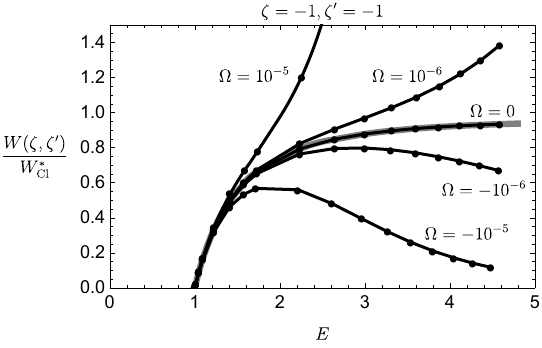} \\
    \includegraphics[width=3in]{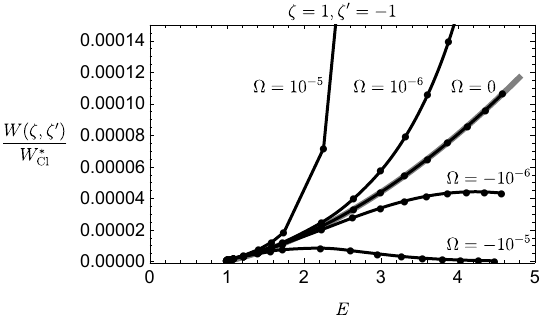} 
    \includegraphics[width=3in]{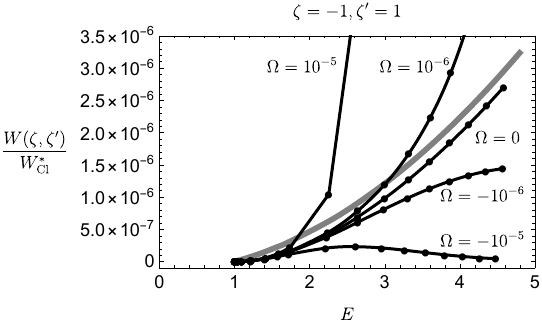} \\
    \caption{The total intensity for $qB = -10^{-3}$ and various $\Omega$. The gray line is the semiclassical result (\ref{eq:SCInt}) with the infrared cutoff (\ref{IRcutoff}). $W_{\text{Cl}}^*$ is given by \eq{Wclass*}.}
    \label{fig:TotIntB3}
\end{figure}

In Figs.~\ref{fig:TotIntB2} and \ref{fig:TotIntB3}, we plot $W(\zeta,\zeta')$ for two values of $|qB|$, integrated over $\theta$ and summed over photon polarizations $h$. Notice the energy scale difference between the two figures, as we have computed with fixed $n$, rather than fixed $E$, see \eq{LL}. We see that in agreement with our expectations $\Omega > 0$ increases the radiation intensity, while $\Omega < 0$ decreases it. Where curves go off the figure, computing the functions $I_{n,n'}(x)$ with acceptable accuracy becomes numerically expensive due to the presence of high-degree polynomials. It was demonstrated in \cite{Buzzegoli:2024nzd} that taking into account the causality effects due to the light-cylinder boundary moderates the steep dependence on the angular velocity. We therefore exhibit only those results that have little dependence on the boundary.

\begin{figure}[ht]
    \centering
    \includegraphics[width=3in]{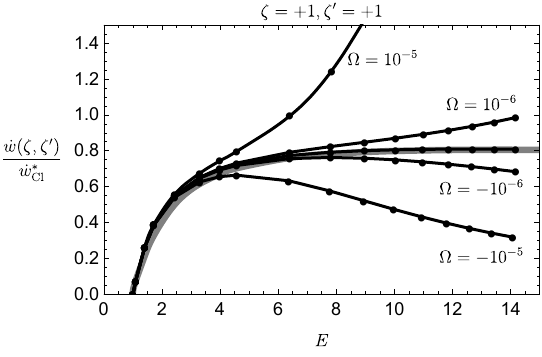} 
    \includegraphics[width=3in]{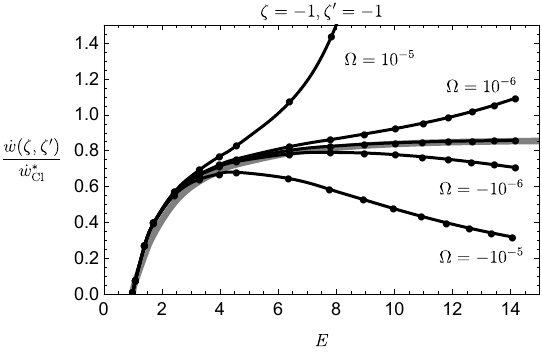} \\
    \includegraphics[width=3in]{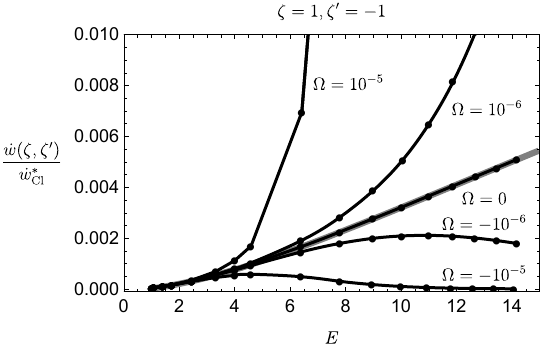} 
    \includegraphics[width=3in]{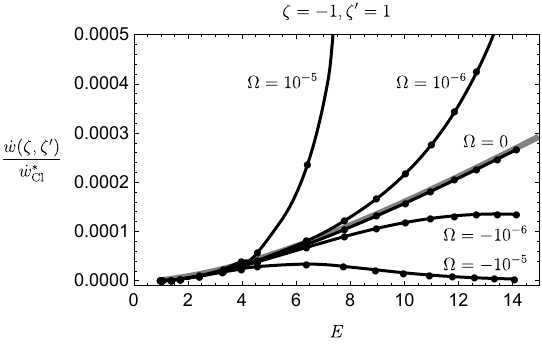} \\
    \caption{The total rate for $qB = -10^{-2}$ and various $\Omega$. The gray line is the semiclassical result \eq{eq:SCRate} with the infrared cutoff \eq{IRcutoff}. $\dot{w}_{\text{Cl}}^*$ is given by \eq{Wclass2*}.}
    \label{fig:TotRateB2}
\end{figure}

\begin{figure}
    \centering
    \includegraphics[width=3in]{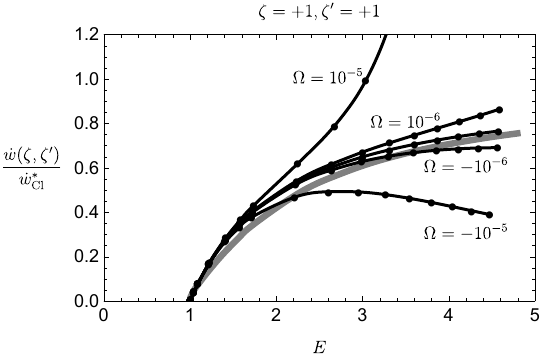} 
    \includegraphics[width=3in]{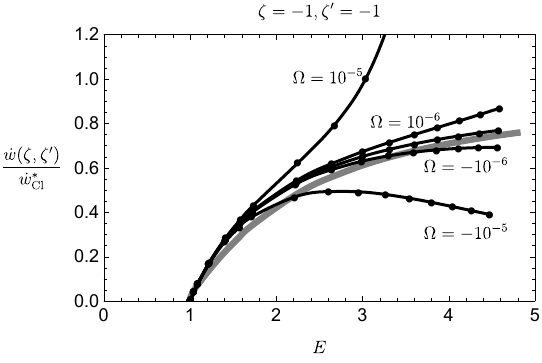} \\
    \includegraphics[width=3in]{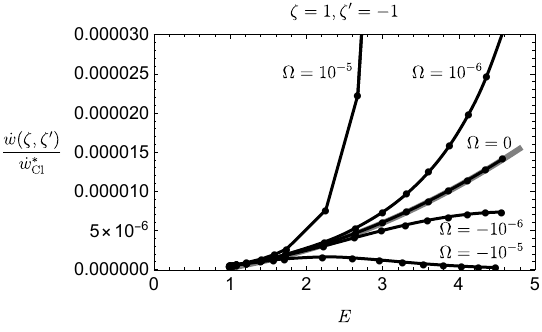} 
    \includegraphics[width=3in]{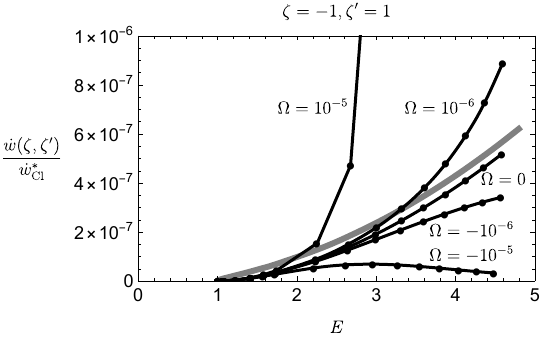} \\
    \caption{The total rate for $qB = -10^{-3}$ and various $\Omega$. The gray line is the semiclassical result \eq{eq:SCRate} with the infrared cutoff \eq{IRcutoff}. $\dot{w}_{\text{Cl}}^*$ is given by \eq{Wclass*}.}
    \label{fig:TotRateB3}
\end{figure}

We plot the total rate, integrated over $\theta$ and summed over $h$, in Figs.~\ref{fig:TotRateB2} and \ref{fig:TotRateB3}. The same trend as in the total intensity is seen, which is not surprising in view of the simple relationship between the intensity and rate, see \eq{eq:Rate} and \eq{eq:RadInt}.

\subsection{Degree of polarization}\label{sec:ST}

The Sokolov--Ternov effect is the self-polarization of charged particles in a constant magnetic field due to the difference in the rate for spin-flip transitions \cite{SokolovAndTernov}. As seen in Figs.~\ref{fig:TotRateB2} and \ref{fig:TotRateB3}, the probability for a spin-up electron to emit a photon and flip its spin is roughly an order of magnitude larger than the probability for a spin-down electron to do the same. We can quantify this difference with the ratio $\eta$, known as the degree of polarization:
\begin{equation}
    \eta = \frac{\dot{w}(\zeta=+1, \zeta'=-1) - \dot{w}(\zeta=-1, \zeta'=+1)}{\dot{w}(\zeta=+1, \zeta'=-1) + \dot{w}(\zeta=-1, \zeta'=+1)}\,.
    \label{eq:eta}
\end{equation}
This is twice the negative expectation value of the $z$-component of the fermion spin. In the high-energy and $\chi \ll 1$ limit, this value approaches the known result \cite{SokolovAndTernov}
\begin{equation}
    \eta_0 = \frac{8\sqrt{3}}{15} = 0.924 ,
\end{equation}
which can be derived from the $\chi \ll 1$ limit of the semiclassical rate.

\begin{figure}
    \centering
    \includegraphics[width=3in]{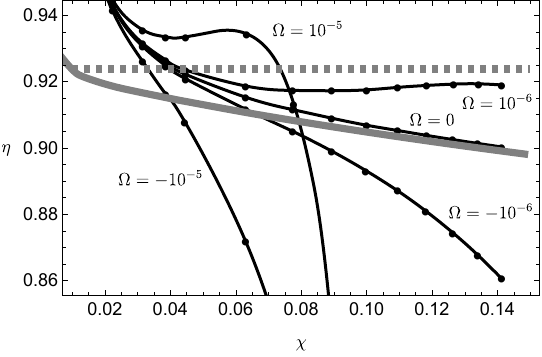}
    \includegraphics[width=3in]{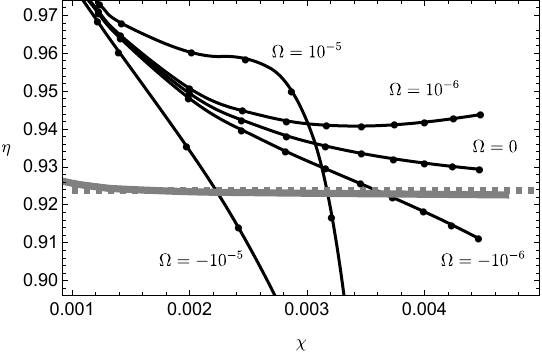}
    \caption{The degree of polarization $\eta$ as a function of $\chi = |qB| Ev$ for $qB = -10^{-2}$ (left) and $qB = -10^{-3}$ (right) for various $\Omega$. The gray lines are two semiclassical results: the dashed line is the 1st order result found in \cite{SokolovAndTernov,Berestetskii:1982qgu}, while the solid line is derived from the result (\ref{eq:SCRate}) with the infrared cutoff (\ref{IRcutoff}).}
    \label{fig:STRatio}
\end{figure}

We plot $\eta$ against the parameter $\chi = |qB| E v$ in Figure \ref{fig:STRatio} for two different fields. First, notice that in the case $|qB| = 10^{-2}$, the full semiclassical result (gray) differs significantly from the value of $\eta_0$. This full result nevertheless agrees qualitatively with the exact numerical calculation for $\Omega = 0$ at larger energies. Contrast this with the case $|qB| = 10^{-3}$, where the two semiclassical results agree, and the exact numerical calculation is approaching the two. The main result of this paper is that the rotation of the medium strongly affects the ratio $\eta$. Especially for the larger values of $|\Omega|$ shown, the difference is quite considerable.

\subsection{Magnetization of plasma by rotation}
\begin{figure}
    \centering
    \includegraphics[width=3in]{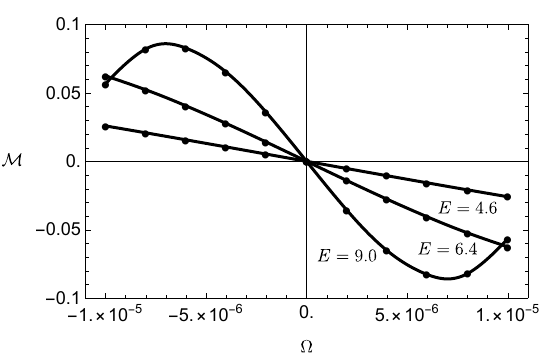}
    \includegraphics[width=3in]{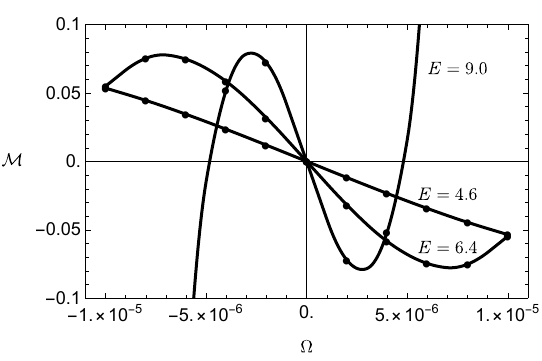}
    \caption{The induced magnetization $\mathcal{M}$ of a rotating plasma due to the Sokolov--Ternov effect at constant energy. The left (right) figure has $|qB| = 10^{-2}$ ($|qB| = 0.5 \times 10^{-2}$).}
    \label{fig:fermion-mag}
\end{figure}

For a given angular velocity and magnetic field the degree of polarization along the symmetry axis depends on the electric charge: as discussed in \cite{Buzzegoli:2023vne}, changing the sign of $q$ is equivalent to changing the sign of $\Omega$ when calculating $W$ or $\dot{w}$. Consider, then, a charge-neutral plasma of particles $f$ of charge $q < 0$ and their antiparticles $\bar{f}$ which rotates with an average vorticity parallel to an external magnetic field. The particles will acquire polarization $\eta_{f}$ in the direction opposite the magnetic field, while the antiparticles will acquire polarization $\eta_{\bar{f}}$ in the same direction as the field. The result in Figure \ref{fig:STRatio} shows that the two species will be unequally polarized: $\eta_f$ will be larger than $\eta_{\bar{f}}$. So, given sufficient time, the plasma will have an overall spin polarization in the direction opposite to the magnetic field. This induces an overall magnetization $\mu N (-\eta_f+\eta_{\bar f})$ in the plasma,  where $\mu$ is the average magnetic moment and $N$ is the number of fermions. The maximal possible magnetization is $\mu N$. The degree of magnetization, i.e.\ the ratio of the magnetization to its maximum possible value, is 
\begin{equation}
    \mathcal{M} = -\eta_{f} + \eta_{\bar{f}} .
\end{equation}
The relative minus sign comes from the definition of $\zeta$, which depends on the sign of $qB$.
$\mathcal{M}$ can also be written as
\begin{equation}
    \mathcal{M} = -\eta_{f}(\Omega) + \eta_{\bar{f}}(\Omega) = -\eta_{f}(\Omega) + \eta_{f}(-\Omega)
    \label{eq:MDef}
\end{equation}
due to the symmetry between sign changes of $qB$ and $\Omega$. We have plotted $\mathcal{M}$ in Fig.~\ref{fig:fermion-mag} for a few fixed energies. We see that for small $\Omega$ and low energy, the induced degree of magnetization $\mathcal{M}$ is linear in $\Omega$. At higher energies, cubic dependence on $\Omega$ emerges. These effects emerge more quickly when $|qB|$ is smaller. The apparent change of sign of the $\partial \mathcal{M}/\partial \Omega$ as the angular velocity increases can be traced back to the sharp dependence of the rate on $\Omega$, which stems from the competition between $\Omega$ and $\omega_B$ explained in the first paragraph of \sec{sec:NumRate}. 

\begin{figure}
    \centering
    \includegraphics[width=3in]{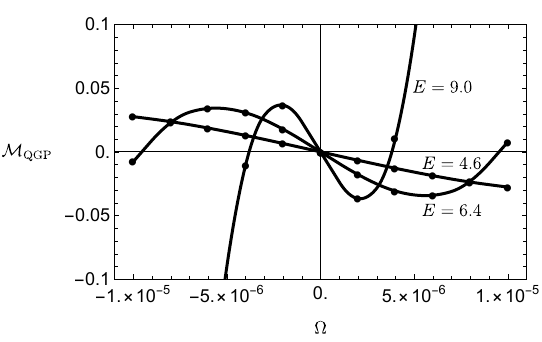}
    \caption{The magnetization $\mathcal{M}$ of quark-gluon plasma due to the Sokolov--Ternov effect at constant energy $E$ (in units of the fermion mass). For each curve, $|eB| = 1.5 \times 10^{-2}$. }
    \label{fig:qgp-mag}
\end{figure}

Now consider a system with many species of fermions, such as quark-gluon plasma. For simplicity, suppose the plasma is composed of $u$ and $d$ quarks and their antiparticles. It was shown in \cite{Tuchin:2010vs} that the maximal degree of polarization $\eta$ in quark-gluon plasma is achieved very rapidly. We can then compute the magnetization induced by the rotation of the plasma in the manner just described, adding the magnetizations from the two fermion species:
\begin{equation}
    \mathcal{M}_\text{QGP} = \eta_{u}(\Omega) - \eta_{\bar{u}}(\Omega) - \eta_{d}(\Omega) + \eta_{\bar{d}}(\Omega) .
\end{equation}
We have been careful in placing the signs to agree with the definition in Eq.~(\ref{eq:MDef}), which assumes the fermion has $q < 0$. We plot the result in Fig.~\ref{fig:qgp-mag} for a few fixed energies.  Evidently quark-gluon plasma may acquire a net magnetization due to its average vorticity.
A more precise phenomenological modeling is beyond the scope of this work.

\section{Conclusions}\label{sec:summary}
We studied the radiative polarization of fermions embedded in a rotating system in a magnetic field. We observed a strong effect of rotation consistent with our previous observations in \cite{Buzzegoli:2022dhw,Buzzegoli:2023vne}. We have only studied a small region of the parameter space, but our conclusions seem to be fairly general. The main difficulty in exploring the parameter space lies in the numerical cost of the calculation: increasing either $n$ or $|\Omega|$ increases the number of transition channels with a significant rate or intensity quickly. A semiclassical approximation taking into account the effects of rotation may alleviate this issue, but the necessary inclusion of a sum or integral over the quantum number $a$ may pose problems (both mathematically and computationally) in that case.

We also note that larger $|\Omega|$ requires careful consideration of the boundary conditions on the light cylinder \cite{Buzzegoli:2024nzd}. We have not taken this into account here, but, as explained in Section \ref{sec:fermion-WF}, by keeping $n+a \ll |qB|/(2\Omega^2)$ we may safely do so. That being said, Figure \ref{fig:STRatio} shows that dramatic behavior can emerge even within this bound.

Rotation has a strong effect on the degree of polarization, which we conjecture could lead to an overall magnetization of a rotating plasma subject to a magnetic field. We have calculated this magnetization, albeit at $T=0$ and with a single initial state. It would be interesting to apply our results to a plasma of fermions at finite temperature and more phenomenologically appropriate parameters, but the numerical cost of such a task pushes it outside the scope of this work. Again, a semiclassical approximation could help, but raises the same issues that we noted above.

\acknowledgments
We thank Matteo Buzzegoli and Nandagopal Vijayakumar for many fruitful discussions of related problems. 
This work  was supported in part by the U.S. Department of Energy under Grant No.\ DE-SC0023692.

\bibliographystyle{apsrev4-1}
\bibliography{biblio}

\appendix
\section{Semiclassical approximation for $\Omega = 0$}
\label{sec:Semiclassical}

At high fermion energies, the rate and intensity in Eqs.~(\ref{eq:Rate}) and (\ref{eq:RadInt}) can be expanded to give a semiclassical approximation. A full treatment of this approximation can be found, for example, in \cite{SokolovAndTernov}. It treats the sum over $n'$ as an integral and converts the Laguerre functions $I_{n,n'}(x)$ into modified Bessel functions $K$ (or, equivalently, Airy functions), which are more easily computed. In the following discussion, we will omit the details of this conversion and skip to the result. We then present a simplified form of the result. We compared this result to our numerical calculations in Section \ref{sec:numerics}.

Following \cite{SokolovAndTernov}, one finds
\begin{gather}
    \dot{w}(\zeta,\zeta') = \dot{w}_{\text{Cl}}\frac{3}{10\pi} \int_0^\infty \frac{dy}{(1 + \xi_0 y)^3} g(y)\,, \\
    W(\zeta,\zeta') = W_{\text{Cl}} \frac{9 \sqrt{3}}{16\pi} \int_0^\infty \frac{y dy}{(1 + \xi_0 y)^4} g(y)\,, \\
    g(y) = \frac{1 + \zeta \zeta'}{2} g^{(1)}(y) + \frac{1 - \zeta \zeta'}{2} g^{(2)}(y) \,,\\
    g^{(1)}(y) = 2 (1 + \xi_0 y) \int_y^\infty K_{5/3}(x) dx + \xi_0^2 y^2 K_{2/3}(y) - \zeta \xi_0 y (2 + \xi_0 y) K_{1/3}(y)\,, \\
    g^{(2)}(y) = \xi_0^2 y^2 (K_{2/3}(y) + \zeta K_{1/3}(y))\, ,
\end{gather}
where $\xi_0 = 3 |qB| E v/ 2= 3\chi/2$ and $y = 2 \omega / (3(E - \omega)\chi$ integrates over $\omega$. Using a recurrence relation of the Bessel $K$ functions
\begin{equation}
    K_{1/3}(y) + K_{5/3}(y) = -2 K_{2/3}'(y)\,,
\end{equation}
and the relationship between the Bessel $K$ and Airy functions,
\begin{gather}
    \Ai(z) = \frac{1}{\pi} \sqrt{\frac{z}{3}} K_{1/3}(\tfrac{2}{3} z^{3/2})\,, \\
    \Ai'(z) = \frac{-1}{\pi} \frac{z}{\sqrt{3}} K_{2/3}(\tfrac{2}{3} z^{3/2})\,,
\end{gather}
 by setting $y = \frac{2}{3} z^{3/2}$, we get
\begin{gather}
    g^{(1)}(y) = -\pi\sqrt{3} \left((2 + \xi_0 y)^2 \frac{1}{z} \Ai'(z) + \zeta \xi_0 y (2 + \xi_0 y) \frac{1}{\sqrt{z}} \Ai(z) + 2 (1 + \xi_0 y) \int_z^\infty \Ai(x) dx\right)\,, \\
    g^{(2)}(y) = -\pi\sqrt{3} \xi_0^2 y^2 \left(\frac{1}{z} \Ai'(z) - \zeta \frac{1}{\sqrt{z}} \Ai(z)\right)\, .
\end{gather}
 The rate and intensity then become
\begin{gather}
    \dot{w}(\zeta,\zeta') = - \dot{w}_{\text{Cl}}\frac{3\sqrt{3}}{10} \int_0^\infty \frac{dz}{(1 + \chi z^{3/2})^3} G(z)\,,
    \label{eq:RateSC1} \\
    W(\zeta,\zeta') = - W_{\text{Cl}} \frac{9}{8} \int_0^\infty \frac{z^{3/2} dz}{(1 + \chi z^{3/2})^4} G(z) \,,
    \label{eq:IntSC1} \\
    G(z) = \frac{1 + \zeta \zeta'}{2} G^{(1)}(z) + \frac{1 - \zeta \zeta'}{2} G^{(2)}(z) = -\frac{\sqrt{z}}{\pi\sqrt{3}} g(y)\,, \\
    G^{(1)}(z) = (2 + \chi z^{3/2})^2 \frac{1}{\sqrt{z}} \Ai'(z) + \zeta \chi z^{3/2} (2 + \chi z^{3/2}) \Ai(z) + 2 (1 + \chi z^{3/2}) \sqrt{z} \int_z^\infty \Ai(x) dx \,,
    \label{eq:G1} \\
    G^{(2)}(z) = \chi^2 z^3 \left(\frac{1}{\sqrt{z}} \Ai'(z) - \zeta \Ai(z)\right) .
    \label{eq:G2}
\end{gather}

These can then be simplified further by integrating by parts. The last term in Eq.~(\ref{eq:G1}) can be made proportional to $\Ai(z)$, and the integral relation
\begin{equation}
    \int z \Ai(z) dz = \Ai'(z)
\end{equation}
can be used to make all terms proportional to $\Ai'(z)$. Due to the difference in form for the rate and intensity, the analyses must be carried out separately, but the calculation is not difficult. The result is
\begin{align}\label{eq:SCRate}
        \dot{w}(\zeta,\zeta') = - \dot{w}_{\text{Cl}} \frac{\sqrt{3}}{20} \int_0^\infty dz \frac{\Ai'(z)}{\sqrt{z}(1 + \chi z^{3/2})^3} \Bigg\{ \frac{1 + \zeta\zeta'}{2} \bigg[2 (10 + 14 \chi z^{3/2} + 7 \chi^2 z^3) \nonumber\\
        - 3 \zeta \chi \frac{2 - 12 \chi z^{3/2} - 5 \chi^2 z^3}{1 + \chi z^{3/2}} \bigg] 
        \nonumber\\
         + \frac{1 - \zeta\zeta'}{2} 3 \chi z^{3/2} \left[2 \chi z^{3/2} + \zeta \chi \frac{7 - 5 \chi z^{3/2}}{1 + \chi z^{3/2}}\right]\Bigg\} \,,
    \end{align}
    \begin{align}    \label{eq:SCInt}
        W(\zeta,\zeta') = - W_{\text{Cl}} \frac{3}{16} \int_0^\infty dz \frac{z \Ai'(z)}{(1 + \chi z^{3/2})^4} \Bigg\{\frac{1 + \zeta\zeta'}{2} \bigg[2 \left(8 + 10 \chi z^{3/2} + 5 \chi^2 z^3\right)\nonumber\\
        - 3 \zeta \chi \frac{8 - 9 \chi z^{3/2} - 5 \chi^2 z^3}{1 + \chi z^{3/2}} \bigg]\nonumber\\
         + \frac{1 - \zeta\zeta'}{2} 3 \chi z^{3/2}\left[2 \chi z^{3/2} + \zeta \chi \frac{7 - 5 \chi z^{3/2}}{1 + \chi z^{3/2}}\right]\Bigg\}\, .
    \end{align}

Expanding to second order in $\chi$ gives the results written in \cite{SokolovAndTernov}.

\section{Infrared cutoff of the photon spectrum}\label{sec:IR}

We found that while the semiclassical equations such as \eq{eq:SCInt} are accurate for the intensity, the same is not true for the rate: the semiclassical formulae \eq{eq:SCRate} derived in \cite{SokolovAndTernov} for processes \emph{without a spin flip} tend to moderately disagree with the exact calculation.

\begin{figure}
    \centering
    \includegraphics[width=3in]{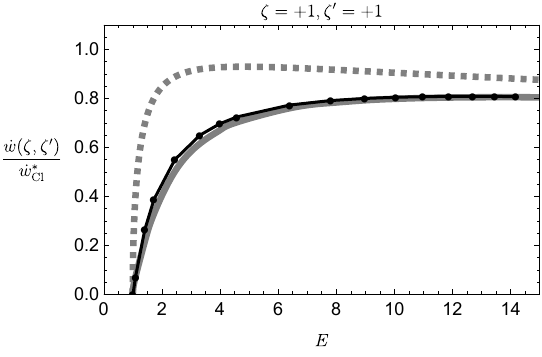}
    \caption{The effect of the cutoff on the rate $\dot{w}$ for $B = 10^{-2}$, $\Omega = 0$, and $\zeta,\zeta' = +1$. The black curve is our numerically calculated result. The solid (dashed) gray curve is the semiclassical result (\ref{eq:SCRate}) (without) the infrared cutoff (\ref{IRcutoff}). $\dot{w}_{\text{Cl}}^*$ is given by \eq{Wclass2*}.}
    \label{fig:CutoffComparison}
\end{figure}

The discrepancy originates in the infrared part of the photon spectrum. When computing the intensity and rate in the semiclassical approximation for $\Omega=0$, one usually ignores that the photon spectrum terminates at a finite energy in the infrared due to the quantized energy levels of the fermion. While this approximation is justifiable for the intensity it is noticeably less accurate for the rate. In particular, we found that taking into account an infrared cutoff on $\omega$ makes the semiclassical result much more accurate for the rate when the spin of the fermion does not flip. The semiclassical expression for the rate has the form
\begin{equation}\label{rate-wkb}
    \dot{w} = \int_{\omega_\text{IR}}^{E-1} \frac{d\dot{w}}{d\omega} d\omega\,
\end{equation}
where $\omega_\text{IR}$ is the smallest possible photon energy allowed by the energy and momentum conservation. In the usual semiclassical expansion, $\omega_{\text{IR}} = 0$. We estimate the cutoff by setting $\Omega=0$ in \eq{omega0} and expanding in powers of the small parameter $|qB|/E^2$, consistent with the semiclassical approximation. The result is
\begin{align}\label{w0appr}
   \omega_0\approx \frac{(n-n')|qB|}{E} \,.
\end{align}
Evidently, the infrared cutoff is 
\begin{align}\label{IRcutoff}
    \omega_\text{IR}=\frac{|qB|}{E} ,
\end{align}
corresponding to a transition between adjacent Landau levels. The same expression can be obtained in a different way, by noting that the radiation peaks in the plane perpendicular to the magnetic field. Thus 
\begin{equation*}
    \omega_0\approx  \sqrt{1 + 2n |qB|} - \sqrt{1 + 2n' |qB| + \omega_0^2} ,
\end{equation*}
whose solution is \eq{w0appr}
\begin{equation}
    \omega_\text{IR} = \frac{(n-n')|qB|}{E} ,
\end{equation}
minimized by $n' = n-1$ (for $n' < n$). This is the smallest photon frequency that can be emitted in the plane of the fermion, where the rate and intensity are greatest. This quantity is small enough that it does not have a significant effect on the intensity or spin-flip rates. Fig.~\ref{fig:CutoffComparison} shows the effect of this cutoff on one of the rates in question. The isolated effect of the cutoff is due to the $\omega^{-2/3}$ divergence of the integrand of \eq{rate-wkb} for transitions without a spin flip as $\omega \to 0$, which is regulated in the intensity by the extra factor of $\omega$ and in the spin-flip rates by their dependence only on larger powers of $\omega$.

\end{document}